\documentclass{PoS}

\def\kHz{{\rm kHz}}
\def\MHz{{\rm MHz}}

\def\m{{\rm m}}

\def\K{{\rm K}}

\def\mJy{{\rm mJy}}
\def\Msun{{\rm M_\odot}}

\title{21cm Forest with the SKA}

\ShortTitle{21cm Forest with the SKA}

\author{Benedetta Ciardi\\
        Max Planck Institute for Astrophysics, Garching, Germany\\
        E-mail: \email{ciardi at mpa-garching.mpg.de}}
\author{Susumu Inoue\\
        Institute for Cosmic Ray Research, University of Tokyo, Tokyo, Japan\\
        E-mail: \email{sinoue at icrr.u-tokyo.ac.jp}}
\author{Katherine J. Mack\\
        University of Melbourne, Melbourne, Australia\\
        E-mail: \email{kmack at unimelb.edu.au}}
\author{Yidong Xu\\
        National Astronomical Observatories, Chinese Academy of Sciences, Beijing, China\\
        E-mail: \email{xuyd at nao.cas.cn}}        
\author{\speaker{Gianni Bernardi}\\
        SKA SA, Cape Town, South Africa\\
        Rhodes University, Grahamstown, South Africa\\
        E-mail: \email{giannibernardi at ska.ac.za}}

\FullConference{
Advancing Astrophysics with the Square Kilometre Array\\
June 8-13, 2014\\
Giardini Naxos, Italy}

\newcommand{\skipthis}[1]{}

\newcommand\apj{ApJ}

\abstract{
An alternative to both the tomography technique and the power spectrum approach is to search for the 21cm forest, that is the 21cm absorption features against high-$z$ radio loud sources caused by the intervening cold neutral intergalactic medum (IGM) and collapsed structures. Although the existence of high-$z$ radio loud sources has not been confirmed yet, SKA-low would be the instrument of choice to find such sources as they are expected to have spectra steeper than their lower-$z$ counterparts. Since the strongest absorption features arise from small scale structures (few tens of physical kpc, or even lower), the 21cm forest can probe the HI density power spectrum on small scales not amenable to measurements by any other means. Also, it can be a unique probe of the heating process and the thermal history of
the early universe, as the signal is strongly dependent on the IGM temperature.
Here we show what SKA1-low could do in terms of detecting the 21cm forest in the redshift range $z \sim 7.5-15$.
}

\begin{document}

\section{Introduction}

An alternative to both the tomography technique and the power spectrum approach is to search for the 21cm forest, that is the 21cm absorption features against high-$z$ radio loud sources caused by the intervening cold neutral IGM and collapsed structures (e.g. Carilli, Gnedin \& Owen 2002; Furlanetto \& Loeb 2002; Furlanetto 2006; Carilli et al. 2007; Xu et al. 2009; Xu, Ferrara \& Chen 2011; Mack \& Wyithe 2012;  Ciardi et al. 2013; Ewall-Wice et al. 2014). In fact the 21cm forest is more than a complement to tomography or power spectrum analysis. Since the strongest absorption features arise from small scale structures, the 21cm forest can probe the HI density power spectrum on small scales not amenable to measurements by any other means (e.g. Shimabukuro et al. 2014). Also, it can be a unique probe of the heating process and the thermal history of the early universe, as the signal is strongly dependent on the IGM temperature.

The photons emitted by a radio loud source at redshift $z_s$ with
frequencies $\nu>\nu_{\rm 21cm}$, will be removed from the source
spectrum with a probability $1-e^{-\tau_{\rm 21cm}}$, absorbed by the neutral hydrogen present along
the line of sight (LOS) at redshift $z=\nu_{\rm 21cm}/\nu(1 + z_s)-1$.  The optical depth $\tau_{\rm 21cm}$ can be written as (e.g. Madau, Meiksin \& Rees 1997; Furlanetto, Oh \& Briggs 2006):
\begin{eqnarray}
\label{eq:tau}
\tau_{\rm 21cm}(z) & = & \frac{3}{32 \pi} \frac{h_p c^3 A_{\rm 21cm}}{k_B \nu_{\rm 21cm}^2}
\frac{x_{\rm HI} n_{\rm H}}{T_s (1+z) (dv_\parallel/dr_\parallel)}, 
\end{eqnarray} 
where $n_{\rm H}$ is the H number density, $x_{\rm HI}$ is the mean neutral hydrogen fraction, 
$T_s$ is the gas spin temperature (which quantifies the relative population
of the two levels of the $^2S_{1/2}$ transition), $A_{\rm 21cm}=2.85 \times 10^{-15}$~s$^{-1}$
is the Einstein coefficient of the transition and 
$dv_\parallel/dr_\parallel$ is the gradient of the proper velocity along the LOS 
(in km~s$^{-1}$), which takes into account also the contribution of the gas peculiar velocity. 
The other symbols appearing in the equation above have the standard meaning adopted in the
literature. 

Analogously to the case of the Ly-$\alpha$ forest, this could result in an average
suppression of the source flux (produced by diffuse neutral hydrogen),
as well as in a series of isolated absorption lines (produced by
overdense clumps of neutral hydrogen), with the strongest absorption
associated with high density, neutral and cold patches of gas.
This suggests that the absorption features due to collapsed
structures with no or very small star formation (to maximize the amount of HI available for absorption), 
such as minihalos or dwarf galaxies
(Furlanetto \& Loeb 2002; Meiksin 2011; Xu, Ferrara \& Chen 2011) would
be easier to detect than those due to the diffuse neutral IGM.
However, this does strongly depend on the feedback effects acting on
such objects. Because of the large uncertainties in the nature and
intensity of high-$z$ feedback effects (for a review see
Ciardi \& Ferrara 2005 and its updated version
arXiv:astro-ph/0409018), it is not straightforward to estimate the
relative importance of the absorption signals from the diffuse IGM and
from collapsed objects.

While gas which has been (even only partially) ionized has a
temperature of $\sim 10^4$~K, gas which has not been reached by
ionizing photons has a temperature which can be even lower that of the
CMB. This neutral gas can be heated by Ly-$\alpha$ or X-ray photons,
thus reducing the optical depth to 21cm. While Ly-$\alpha$ heating is
not extremely efficient, heating due to X-ray photons could easily
suppress the otherwise present absorption features
(e.g. Mack \& Wyithe 2012; Ciardi et al. 2013). This seems to
suggest that with observations of the 21cm forest it would be possible to
discriminate between different IGM reheating histories, in particular
if a high energy component in the ionising spectrum were present
(Ewall-Wice et al. 2013).

\section{Observed spectra}

In this Section we describe the process used to simulate an observed spectrum. 
The simulated absorption spectrum, $S_{\rm abs}$, is calculated from a full 3D radiative
transfer simulation of IGM reionization which resolves scales of $\sim
1$~kHz (corresponding to $\sim$50 kpc comoving; Ciardi et al. 2012; Ciardi et al. 2013). In the simulation used here, the contribution 
to ionization and heating from x-rays and/or Ly-$\alpha$ photons is not included (although see Ciardi et al. 2013 for
examples of spectra which include such effects), i.e. the gas which is not reached by UV ionizing photons remains cold.

The instrumental effects are estimated using the pipeline of the LOFAR telescope. More specifically,
the equation giving the observed visibilities is:

\begin{equation}
{V_\nu }\left( {\bf{u}} \right) = \sum\limits_i^{{N_{sources}}} {{I_\nu }({\bf{s}})} 
{e^{ - 2\pi i{\bf{u}} \cdot {\bf{s}}}} + n_s,
\end{equation}
where $\mathbf{u}=(u, v, w)$ are the  coordinates of a given baseline at a certain 
time $t$, ${I_\nu }$ is the observed source intensity, $\mathbf{s}=(l,m, n)$ is 
a vector representing the direction cosines for a given source direction and $n_s$ 
represents additive noise.  The noise is given by the radiometer equation:

\begin{equation}
n_s = \frac{1}{{{\eta_s}}}\frac{{SEFD}}{{\sqrt {2{t_{{\mathop{\rm int}} }} \Delta \nu} }},
\label{eq:noise}
\end{equation}
where $\eta_s$ is the system efficiency, $\Delta\nu$ is the bandwidth and $t_{int}$ is 
the integration time.
Because of our experience with the LOFAR telescope, we will make predictions based on its characteristics, and we will then scale the noise to match the one expected from SKA.
We thus assume that  $\eta_s=0.5$
and $N_{st}=48$.  The system equivalent flux density is given by:

\begin{equation}
SEFD = \frac{{2{\kappa _B}{T_{\rm sys}}}}{{N_{\rm dip}{\eta _\alpha }{A_{\rm eff}}}},
\end{equation}
where $\kappa _B$ is Boltzmann's constant, $A_{\rm
eff}=min(\frac{\lambda2}{3}, 1.5626)$ is the effective area of each
dipole in the dense and sparse array regimes respectively, $N_{\rm
dip}$ is the number of dipoles per station (24 tiles times 16 dipoles
per tile for a LOFAR core station) and $\eta_{\alpha}$ is the
dipole efficiency which we assume to be 1. The system noise $T_{\rm
sys}$ has two contributions: {\it (i)} from the electronics and {\it
(ii)} from the sky. We assume that the sky has a spectral index of
-2.55, obtaining $T_{\rm sys}=[140+60(\nu/150 \; {\rm MHz})^{-2.55}]$~K.
The complete Fourier plane sampling can be done by evaluating the
above equation for every set of baseline coordinates. The predicted
visibilities are then gridded and transformed via inverse Fourier
transforms in order to obtain the dirty images.
For the purpose of the 21~cm forest, fine spectral resolution ($\sim$1~kHz, see later for further discussion) is
needed, which means that the spectra need to be predicted for a large
number of channels ($\sim 10\, 000$). After assembling the full image cube, the LOS
spectrum is extracted.
In principle, the effect of
the Point Spread Function (PSF) side lobes running through the source
of interest can be taken into account by including more sources at
different positions, with or without absorption features. Here, we have ignored the effect of side lobe noise.
In fact, as the SKA will have a very dense uv coverage and the expected 21cm absorption lines will be narrower than a few tens of kHz, we expect the side lobe noise to play a marginal role.                

\begin{figure}[t!]
\centering
\includegraphics[width=1\textwidth]{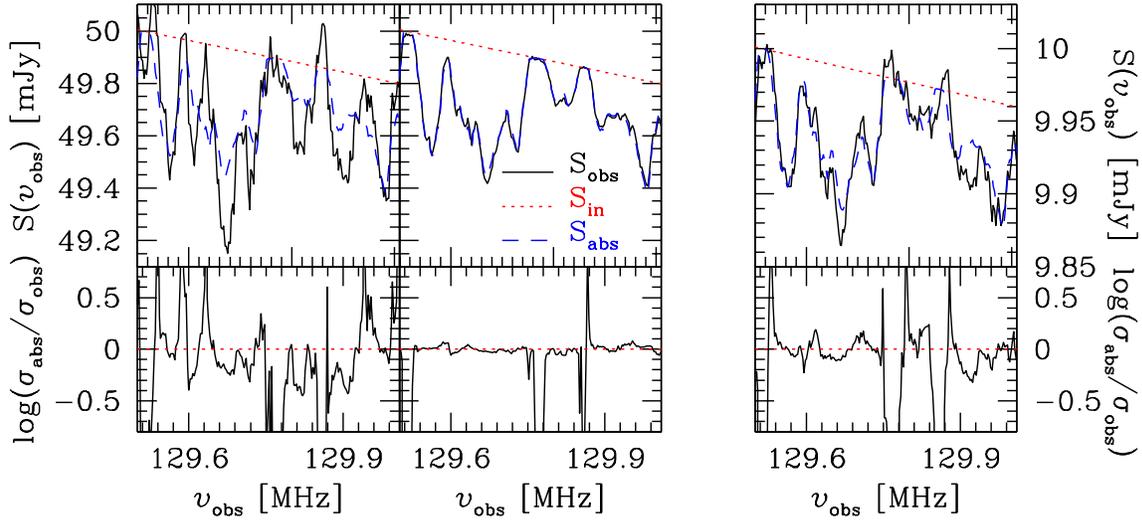}
\caption{\small {\it Upper panels:} Spectrum of a radio source
  positioned at $z=10$ ($\nu \sim 129$~MHz), with a power-law index
  $\alpha=1.05$ and a flux density $J=50$~mJy (left hand panels) and 10~mJy (right hand panel).  The red dotted lines
  refer to the instrinsic spectrum of the radio source, $S_{\rm in}$;
  the blue dashed lines to the simulated spectrum for 21cm absorption,
  $S_{\rm abs}$ (in a universe where neutral regions remain cold); and
  the black solid lines to the spectrum for 21cm absorption as it
  would be seen with an observation time
  $t_\mathrm{int}=1000$~h and a frequency resolution $\Delta
  \nu=10$~kHz. The first panel to the left corresponds to a case with the LOFAR noise, while the other two panels have 1/10th of the LOFAR noise, roughly expected for SKA1-low. 
   {\it Lower panels:} The ratio $\sigma_{\rm abs}/\sigma_{\rm obs}$ corresponding to the upper panels. }
\label{fig:21cmforest_1}
\end{figure}

Figure~\ref{fig:21cmforest_1} shows the 21cm absorption spectrum
due to the diffuse IGM along a random LOS for a bright radio source at $z=10$ (i.e. $\nu
\sim 129$~MHz). For an easy comparison to existing work on the 21cm forest (e.g. Carilli, Gnedin \& Owen 2002; Mack \& Wyithe 2012;  
Ciardi et al. 2013), the intrinsic radio source spectrum, $S_{\rm in}$, is
assumed to be similar to Cygnus A, with a power-law with index
$\alpha=1.05$ and a flux density $J=50$~mJy and 10~mJy.  The simulated absorption
spectrum, $S_{\rm abs}$, is calculated from the simulations mentioned above.
The observed spectrum,
$S_{\rm obs}$, is calculated assuming an observation time
$t_\mathrm{int}=1000$~h with the LOFAR and SKA1-low telescopes and a bandwidth
$\Delta \nu=10$~kHz.  A clear absorption signal is observed. 
This is more evident in the lower
panels of Figure~\ref{fig:21cmforest_1}, which show the quantity
$\sigma_{\rm abs}/\sigma_{\rm obs}$, where $\sigma_i=S_i-S_{\rm in}$
and $i$=abs, obs.
As already mentioned above, the inclusion of Ly-$\alpha$ or x-ray heating could suppress or reduce the absoprtion features, with the extent of the effect being highly dependent on the source model (see e.g. Mack \& Wyithe 2012; Ciardi et al. 2013).

\begin{figure}
\centering
\includegraphics[width=1.\textwidth]{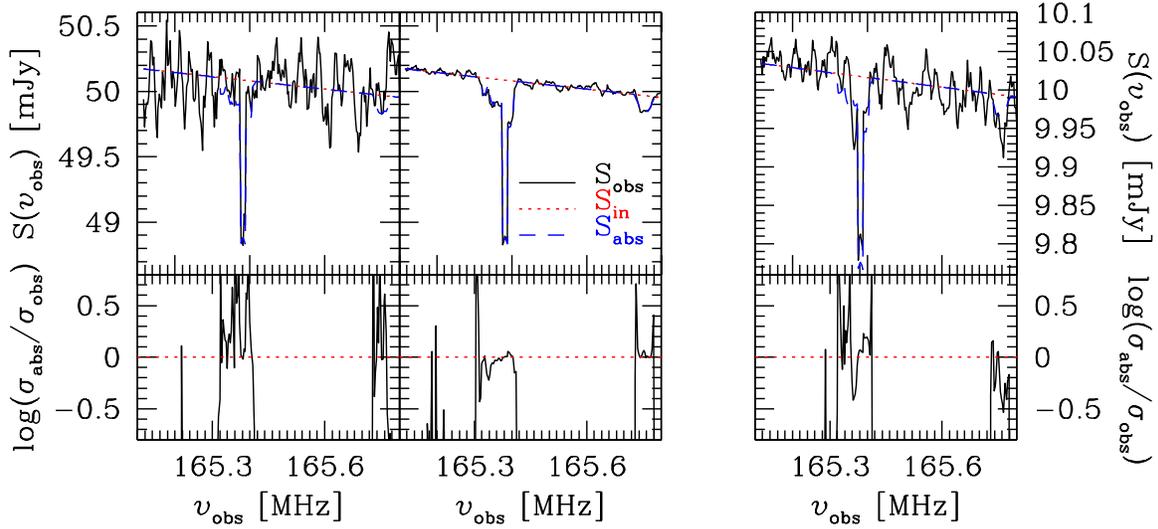}
\caption{\small {\it Upper panels:} Spectrum of a radio source
  positioned at $z=7.6$ ($\nu \sim 165$~MHz), with a power-law index
  $\alpha=1.05$ and a flux density $J=50$~mJy (left hand panels) and 10~mJy (right hand panel).  The red dotted lines
  refer to the instrinsic spectrum of the radio source, $S_{\rm in}$;
  the blue dashed lines to the simulated spectrum for 21cm absorption,
  $S_{\rm abs}$ (in a universe where neutral regions remain cold); and
  the black solid lines to the spectrum for 21cm absorption as it
  would be seen with an observation time
  $t_\mathrm{int}=1000$~h and a frequency resolution $\Delta
  \nu=5$~kHz. The first panel to the left corresponds to a case with the LOFAR noise, while the other two panels have 1/10th of the LOFAR noise, roughly expected for SKA1-low. 
   {\it Lower panels:} The ratio $\sigma_{\rm abs}/\sigma_{\rm obs}$ corresponding to the upper panels. }\label{fig:21cmforest_2}
\end{figure}

Very strong absorption features could be easily detected also at lower redshift, when most of the IGM is in a highly ionization state, if
we were lucky enough to intercept high density cold pockets
of gas  (with $\tau_{\rm 21cm}>0.1$; these cells are found in
$\sim$ 0.1\% of the LOS in the simulation), as shown in Figure~\ref{fig:21cmforest_2}.

\begin{figure}
\centering
\includegraphics[width=1\textwidth]{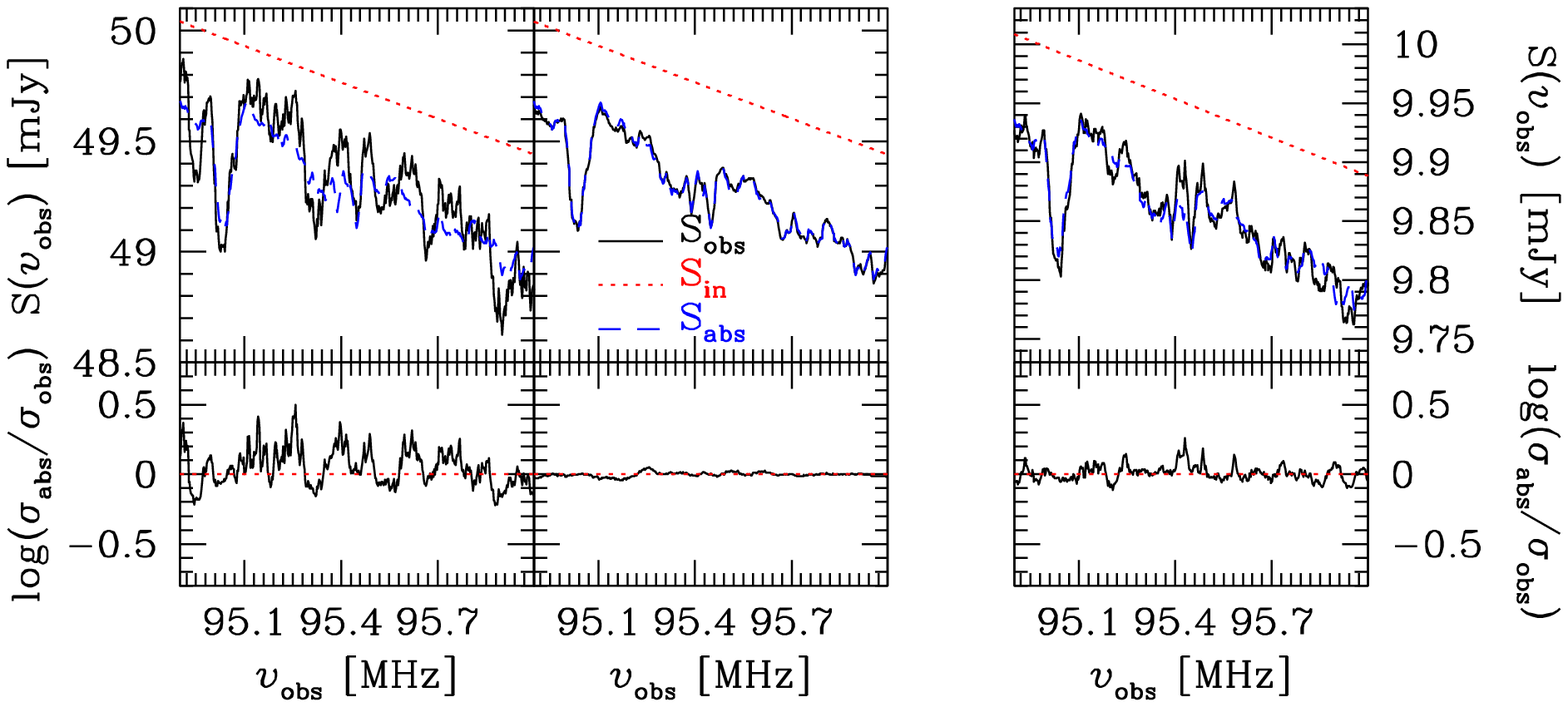}
\caption{\small {\it Upper panels:} Spectrum of a radio source
  positioned at $z=14$ ($\nu \sim 95$~MHz), with a power-law index
  $\alpha=1.05$ and a flux density $J=50$~mJy (left hand panels) and 10~mJy (right hand panel).  The red dotted lines
  refer to the instrinsic spectrum of the radio source, $S_{\rm in}$;
  the blue dashed lines to the simulated spectrum for 21cm absorption,
  $S_{\rm abs}$ (in a universe where neutral regions remain cold); and
  the black solid lines to the spectrum for 21cm absorption as it
  would be seen with an observation time
  $t_\mathrm{int}=1000$~h and a frequency resolution $\Delta
  \nu=20$~kHz. The first panel to the left corresponds to a case with the LOFAR noise, while the other two panels have 1/10th of the LOFAR noise, roughly expected for SKA1-low. 
   {\it Lower panels:} The ratio $\sigma_{\rm abs}/\sigma_{\rm obs}$ corresponding to the upper panels. }
\label{fig:21cmforest_3}
\end{figure}

Moving towards higher redshift, when most of the gas in the IGM is still neutral and relatively cold, would offer the chance of detecting a stronger average absorption (rather than the single absorption features observed at lower redshift). If a radio source with characteristics similar to the ones described above were found, SKA1-low would easily detect the global absorption as shown in Figure~\ref{fig:21cmforest_3}, although it would not be straightforward to distinguish whether the suppression of the source flux were due to the intervining neutral IGM or an intrinsically lower flux.

\section{Challenges}

The most challenging aspect of the detection of a 21cm forest remains
the existence of high-$z$ radio loud sources. Although a QSO has been
detected at $z=7.085$ (Mortlock et al. 2011), it is radio quiet (Momjian et al. 2014), and the existence of
even higher redshift quasars is uncertain. 
To observe the absorption features in the spectrum, this
has to be observed with a certain
precision, which depends on the brightness of the quasar and the sensitivity of the
instruments. The minimum detectable flux density of an interferometer can be
written as:
\begin{equation}
\Delta S_{\rm min}\, =\, \frac{2\, \kappa_B \, T_{\rm sys}}{A\sqrt{\Delta\nu \,t_{int}}}\, \frac{S}{N},
\end{equation}
where $A$ is the collecting area of the array and $S/N$ is the signal-to-noise ratio.
As one may not be able to distinguish whether the flux decrement were
due to the diffuse IGM or an intrinsically lower flux, we will probably only
detect the additional absorptions with respect to the absorption by the IGM.
Therefore, the minimum flux
density of the background source required to observe the absorption
lines is:
\begin{eqnarray}
S_{\rm min}&=&10.3 \mJy \left(\frac{S/N}{5}\right)\left(
\frac{0.01}{\displaystyle e^{-\tau_{\rm IGM}} - e^{-\tau}}\right)
\left(\frac{5 \kHz}{\Delta\nu}\right)^{1/2}
\left(\frac{1000\,\m^2\K^{-1}}{A/T_{\rm
sys}}\right) \left(\frac{1000\, {\rm hr}}{t_{\rm int}}\right)^{1/2}.
\end{eqnarray}
The SKA1-low will have $A/T_{\rm sys} \sim 1000 \m^2\K^{-1}$, 
and we expect $A/T_{\rm sys} \sim 4000 \m^2\K^{-1}$ for SKA2-low.

\begin{figure}
\centering
\includegraphics[width=1\textwidth]{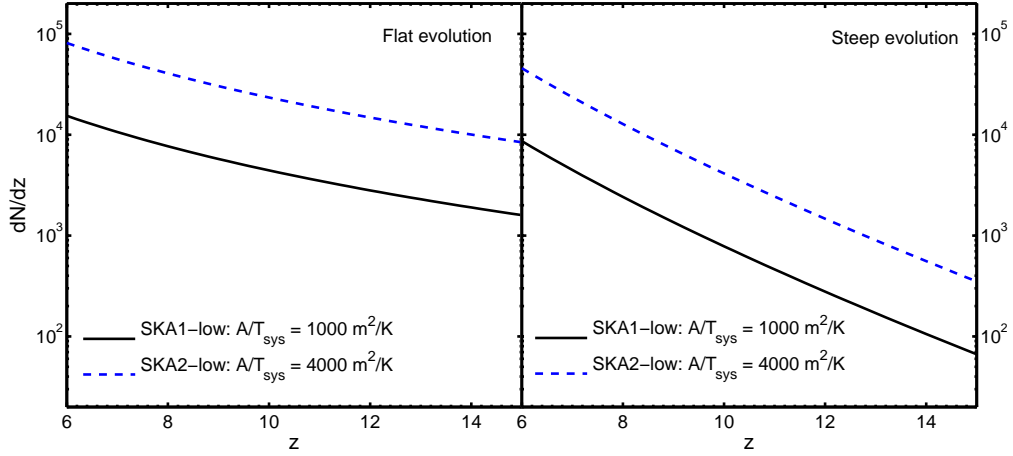}
\caption{The number of quasars in the whole sky that
can be used to
detect signals with $e^{-\tau_{\rm IGM}} - e^{-\tau} \ge 0.01$ per redshift interval. 
The sensitivity of the radio array is taken to be $A/T_{\rm sys} = 1000$ $\m^2\K^{-1}$ (black solid lines) and
$4000$ $\m^2\K^{-1}$ (blue dashed). The integration time is assumed to be 1000 hours.
{\it Left panel}: the number of qualified quasars in the flat evolution
model. {\it Right panel}: the number of qualified quasars in the steep evolution model.
}
\label{fig:21cmforest_4}
\end{figure}

Extrapolating the observed number density of radio sources at $z=4$ (Jarvis et al. 2001)
to higher redshift and lower luminosity,
one can calculate the number of quasars with flux density
at the observed frequency $\nu_{obs} = 1420.4/(1+z)\,\MHz$ larger than the
lower limit described above, for a flat evolution model and a steep evolution model 
respectively (Xu et al. 2009). Using the planned SKA sensitivities, the predicted
numbers of qualified radio sources are plotted in Fig.\ref{fig:21cmforest_4}.

The predicted number of radio sources which can be used for 21cm forest studies in the whole
sky per unit redshift at $z=10$ varies in the range $8\times10^2 - 2\times10^4$. This expected 
number, which is based on observations at redshift lower than reionization,
is very uncertain. The exact number 
depends on the model adopted for the luminosity function of such
sources and the instrumental characteristics
(e.g. Carilli, Gnedin \& Owen 2002; Xu et al. 2009), making such a
detection an extremely challenging task. 
As a reference, the estimated quasar number per unit redshift at $z=10$ for the 3-tiered survey is reported in Table~\ref{tab:qso}.

\begin{table}
\begin{tabular}{|l|l|l|l|l|}
\hline
  $dN/dz$ &  Model & $\Omega$ [sqd] & $t_{\rm int}$ [h] & $A/T_{\rm sys}$ $[\m^2\K^{-1}]$ \\ \hline
11  &  Flat  &  100   &  1000 &  1000 \\ \hline
27  &  Flat  &  1000  &  100  &  1000 \\ \hline
68  &  Flat  &  10000 &  10   &  1000 \\ \hline
57  &  Flat  &  100   &  1000 &  4000 \\ \hline
142 &  Flat  &  1000  &  100  &  4000 \\ \hline
358 &  Flat  &  10000 &  10   &  4000 \\ \hline
2   &  Steep &  100   &  1000 &  1000 \\ \hline
5   &  Steep &  1000  &  100  &  1000 \\ \hline
12  &  Steep &  10000 &  10   &  1000 \\ \hline
10  &  Steep &  100   &  1000 &  4000 \\ \hline
25  &  Steep &  1000  &  100  &  4000 \\ \hline
63  &  Steep &  10000 &  10   &  4000 \\ \hline
\end{tabular}
\caption{Estimated quasar number at $z=10$ for the 3-tiered survey. The columns refer to: number per unit redshift, source model,
survey area, observation time, $A/T_{\rm sys}$.}
\label{tab:qso}
\end{table}

If a sufficiently bright radio loud quasar were found beyond the redshift of reionization,
then the absorption lines generated from early non-linear structures could be
easily detected, as the spectral resolution of the SKA will not be a problem.
Assuming the minimum halo mass hosting cold neutral gas to be $10^6 \Msun$,
about one absorption line in every $8.4\kHz$ is expected at redshift $z\sim 10$.
The absorption line density is lower for lower redshift and higher minimum mass.
On the other hand, the line width from non-linear structures ranges mostly from 
$\sim 1\kHz$ to $\sim 5\kHz$. Given that the SKA1-low will have a spectral resolution 
of $1\kHz$, the line counting is feasible as long as sufficiently bright radio sources are
available at high redshift. Especially, if one stacks together several lines to get an 
average profile, it will hopefully reveal the physical status of the early non-linear structures.

\begin{figure}
\centering
\includegraphics[width=0.8\textwidth]{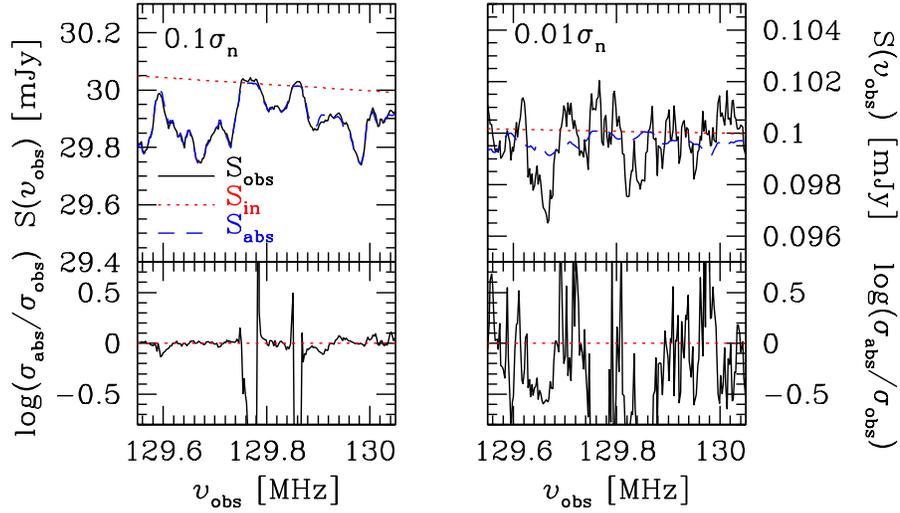}
\caption{\small {\it Upper panels:} Spectrum of a GRB              
  positioned at $z=10$ ($\nu \sim 129$~MHz), with a power-law index
  $\alpha=0.5$ and a flux density $J=30$~mJy (GRB from Pop~III stars; left hand panel) and 0.1~mJy (GRB from Pop~II stars; right hand panel).  The red dotted lines
  refer to the instrinsic spectrum of the source, $S_{\rm in}$;
  the blue dashed lines to the simulated spectrum for 21cm absorption,
  $S_{\rm abs}$ (in a universe where neutral regions remain cold); and
  the black solid lines to the spectrum for 21cm absorption as it
  would be seen with an observation time
  $t_\mathrm{int}=1000$~h and a frequency resolution $\Delta
  \nu=10$~kHz. The panel to the left (right) corresponds to a case with 1/10th (1/100th) of the LOFAR noise, roughly expected for SKA1-low (SKA2-low).
   {\it Lower panels:} The ratio $\sigma_{\rm abs}/\sigma_{\rm obs}$ corresponding to the upper panels. }
\label{fig:21cmforest_5}
\end{figure}

An alternative possibility is the radio afterglows of certain types of gamma-ray bursts (GRBs).
GRBs have already been observed up to $z \sim 8-9$, and it is plausible that they occur up to the earliest epochs
of star formation in the universe at $z \sim 20$ or higher. In addition, they have a simple power-law spectrum
at low frequencies, making the signal extraction easy.
However, if such GRBs are similar to those seen at lower redshifts, their radio afterglows
are not expected to be bright enough at the relevant observed frequencies $\nu_{\rm obs} \lesssim$ 100 MHz
due to strong synchrotron self-absorption (Ioka \& Meszaros 2005; Inoue, Omukai \& Ciardi 2007).
On the other hand, it has been recently proposed that GRBs arising from Population (Pop) III stars
forming in metal-free or very metal poor environments may be much more energetic compared to ordinary GRBs,
leading to blastwaves expanding to much larger radii, and consequently much brighter
low-frequency radio afterglows, exceeding tens of mJy (Toma, Sakamoto \& M{\'e}sz{\'a}ros 2011).
This occurs over timescales of up to $\sim 1000$ yr, making them virtually steady radio sources,
differently from more standard, lower redshift GRBs which, on the other hand, offer relatively short integration times  and a limited spectral length between the location of the GRB and the end of reionization (e.g. Xu et al. 2011).
In Figure~\ref{fig:21cmforest_5} we show an example of absorption spectra in high-$z$ GRBs, from both Pop~III and Pop~II stars (Ciardi et al., in prep).

If the rate of Pop III GRBs with sufficiently bright radio emission is 0.1 yr$^{-1}$ or roughly $10^{-4}$ of all GRBs,
one may expect $\sim 100$ such sources all sky at a given time.
A practical question that remains is how we can observationally identify such sources.
One possibility of using GRB afterglows as the background radio source is the 
broad-band observation, measuring the mean flux decrement in each band
without resolving individual absorption lines (Xu et al. 2011).

\section{Discussion and Summary}

As explained above, absorption features due to small collapsed objects
can be much stronger than those due to the diffuse neutral IGM. Since
their cross-sections are small, the best conditions for detecting them
would be when Ly-$\alpha$ coupling pushes the spin temperature in
their lower density outskirts to the gas temperature before these
regions have been affected by any heating (see Fig.~22 in Meiksin 2011),
conditions expected above $z\sim10$. However, even after heating has started to suppress the 21cm
absorption signal, some weak features due to collapsed structures may
remain. 
Interestingly, even when it may not be possible to detect these weak features individually, the presence of absorption may be detected statistically. Weak absorption would produce an increase in the variance of brightness fluctuations, as an addition to the telescope noise, resulting in an apparently noisier spectrum blue-ward of the 21cm transition (see Fig. 29 in Meiksin 2011; Fig. 8 in Mack \& Wyithe 2012).
The possibility to have a statistical detection of the 21cm forest (rather than the detection of single absorption features), 
is intriguing as the signature of the forest would be observed in the 21cm power spectrum (see also Ewall-Wice et al. 2013). By integrating the signal from many high redshift sources within the field of view, would reduce the sensitivity requirements of the instrument. On the other hand, the main problem of the paucity of high-$z$ radio sources would persist.

The attempt to detect the 21cm forest has the potential to provide unprecedented information about the large-scale evolution of the intergalactic medium as well as the growth of small-scale structures. Different information can be gleaned depending on the nature of the observation. An attempt to detect the presence of the 21cm forest via a statistical analysis of the power spectrum (Ewall-Wice et al. 2013) will allow us to put constraints on the presence of high-redshift radio-loud sources at high wavenumbers ($k \gtrsim$ 0.5 Mpc$^{-1}$), and to study the IGM thermal history at low wavenumbers ($k \lesssim$ 0.1 Mpc$^{-1}$). Assuming high-redshift radio-loud sources are identified, and we are able to observe them directly in targeted observations, detailed study of the source spectra can reveal the mean thermal properties of the IGM via the flux decrement (Furlanetto 2006) or variance in the flux (Carilli et al. 2004; Mack \& Wyithe 2012). If individual absorption features are identified in the spectrum, this could allow us to put constraints on the clumping factor at early times (Xu, Ferrarra \& Chen 2011) and to map structure along the line of sight (IGM and/or the first collapsed structures and mini-halos), thus significantly extending our understanding of the process of reionization and the hierarchical growth of structure in the Universe.

Finally, we would like to comment on the fact that as observations of the 21~cm forest require very high frequency resolution, they are demanding both in terms of data storage and processing. However, when a bright continuum radio source is identified after an all-sky survey or deep EoR observations, it is possible to significantly average in time after phasing up to the source. The maximum amount of data compression will be dictated by the need to identify and subtract bright, far away sources that may generate sidelobe noise in the measurement of the 21~cm forest. 
However, if bright sources are measured a priori from a sky survey they can be subtracted from the high time resolution visibilities and, afterwards, data can be easily compressed by an order of magnitude or more.

\end{document}